\documentclass[aps,prd,twocolumn,superscriptaddress,showpacs]{revtex4}
\pdfoutput=1
\usepackage{amsmath}
\usepackage{graphicx}

\newcommand{\bastar}{\begin{eqnarray*}}
\newcommand{\eastar}{\end{eqnarray*}}
\newskip\humongous \humongous=0pt plus 1000pt minus 1000pt

\newif\ifdtup

\relax
\newcommand{\W}{{\vec W}}

\newcommand{\hr}{\hat r}

\newcommand{\bea}{\begin{eqnarray}}
\newcommand{\eea}{\end{eqnarray}}

\newcommand{\Int}{\displaystyle\int}
\newcommand{\A}{{\vec A}}
\newcommand{\C}{{\vec C}}
\newcommand{\F}{{\vec F}}
\newcommand{\mn}{{\mu\nu}}

\newcommand{\nn}{\nonumber}
\newcommand{\pro}{\partial}

\begin{document}
\title{Mass of the Electroweak Monopole}
\author{Y. M. Cho}
\email{ymcho7@konkuk.ac.kr}
\affiliation{Administration Building 310-4, Konkuk University, 
Seoul 143-701, Korea}
\affiliation{School of Physics and Astronomy,
Seoul National University, Seoul 151-747, Korea}
\author{Kyoungtae Kimm}
\affiliation{Faculty of Liberal Education,
Seoul National University, Seoul 151-747, Korea}
\author{J. H. Yoon}
\affiliation{Department of Physics, 
Konkuk University, Seoul 143-701, Korea}

\begin{abstract}
We present two independent methods to estimate the mass 
of the electroweak monopole. Our result strongly implies 
the existence of a genuine electroweak monopole of mass 
around 4 to 10 TeV, which could be detected by MoEDAL at 
present LHC. We emphasize that the discovery of the the 
electroweak monopole should be the final test of the standard 
model.  
\end{abstract}
\maketitle

The recent ``discovery" of Higgs particle at LHC has reconfirmed 
that the standard model describes the real world \cite{LHC}. 
Indeed it has been interpreted that the standard model has passed 
the ``final" test with the discovery. But we emphasize tha the 
final test of the standard model should come from the discovery 
of the electroweak (``Cho-Maison") monopole, because the theory 
predicts the electroweak monopole \cite{plb97,yang}. It has the 
monopole topology, and naturally accommodates the Cho-Maison 
monopole as the electroweak generalization of the Dirac's 
monopole \cite{dirac}. 

Ever since Dirac predicted the existence of the monopole, the 
monopole has been an obsession \cite{wu,prl80,thooft,julia,dokos}. 
The Abelian monopole has been generalized to the non-Abelian 
monopoles by Wu and Yang \cite{wu,prl80} who showed that the pure 
$SU(2)$ gauge theory allows a point-like monopole, and by 't Hooft 
and Polyakov \cite{thooft,julia} who have constructed a finite energy 
monopole solution in Georgi-Glashow model as a topological soliton. 
Moreover, the monopole in grand unification has been constructed by
Dokos and Tomaras \cite{dokos}.

But it has been asserted that the Weinberg-Salam model has 
no topological monopole of physical interest \cite{vach}. 
The basis for this ``non-existence theorem'' is that the 
quotient space $SU(2) \times U(1)/U(1)_{\rm em}$ allows no 
non-trivial second homotopy which can accommodate the monopole.  

This claim, however, is unfounded \cite{plb97,yang}. This 
is because the Weinberg-Salam model, with the hypercharge 
$U(1)_Y$, could be viewed as a gauged $CP^1$ model in which 
the (normalized) Higgs doublet plays the role of the $CP^1$ 
field. So, if the standard model is correct, the electroweak 
monopole must exist. This makes the experimental detection 
of the electroweak monopole an urgent matter. 

Fortunately the latest MoEDAL detector (``The Magnificient 
Seventh") at LHC is actively searching for such monopole \cite{pin}. 
To help the experiment discover the monopole, however, we need 
a theoretical estimate of the monopole mass. {\it The purpose of 
this Letter is to estimate the mass of the electroweak monopole 
to be around 4 to 10 TeV.} 

The importance of the electroweak monopole is twofold. First, 
it is the straightforward and natural generalization of the 
Dirac monopole to the electroweak theory which is unavoidable 
when the electrodynamics is unified to the electroweak theory. 
This means that the monopole which should exist in the real 
world is not likely to be the Dirac monopole but this one. 

Second, unlike the Dirac monopole which is optional, the 
electroweak monopole must exist because the standard model 
has the monopole topology. This means that the final test 
of the standard model is not the discovery of the Higgs particle, 
but the confirmation of the electroweak monopole. Indeed the 
discovery of the electroweak monopole should be regarded as 
the topological test of the standard model which has never 
been done before.

The Cho-Maison monopole may be viewed as a hybrid between the 
Dirac monopole and the 'tHooft-Polyakov monopole, because it 
has a $U(1)$ point singularity at the center even though the 
$SU(2)$ part is completely regular. Consequently it carries an 
infinite energy at the classical level, so that the mass of the 
monopole can be arbitrary. {\em A priori} there is nothing wrong 
with this, but nevertheless one may wonder whether one can estimate 
the mass of the electroweak monopole. In the following we provide 
three independent methods to estimate the monopole mass.

Consider the standard Weinberg-Salam model,
\begin{gather}
{\cal L} =-|D_{\mu} \phi|^2 -\frac{\lambda}{2}\big(|\phi|^2
-\frac{\mu^2}{\lambda}\big)^2-\frac{1}{4}\F_{\mu\nu}^2 
-\frac{1}{4}G_{\mu\nu}^2  , \nn \\
D_{\mu} \phi = \big(\partial_{\mu} 
  -i\frac{g}{2} \vec \tau \cdot \A_{\mu}
  -i\frac{g'}{2} B_{\mu}\big) \phi,
\label{lag1}
\end{gather}
where $\phi$ is the Higgs doublet, $\F_{\mu\nu}$, $\A_{\mu}$  
and $G_{\mu\nu}$, $B_{\mu}$ are the gauge fields of $SU(2)$ 
and $U(1)_Y$. Now choose the ansatz in the spherical coordinates 
$(t,r,\theta,\varphi)$ 
\begin{gather}
\phi=\frac{1}{\sqrt{2}}\rho(r)\xi(\theta,\varphi),
~~~\xi =i\left(\begin{array}{cc}
\sin (\theta/2)\,\, e^{-i\varphi}\\
- \cos(\theta/2) \end{array} \right), \nn\\
\A_{\mu}= \frac{1}{g} A(r) \partial_{\mu} t~\hat r
+\frac{1}{g}(f(r)-1)~\hat r \times \partial_{\mu} \hat r, \nn\\
\hat r=-\xi^{\dagger} \vec \tau \xi,  \nn\\
B_{\mu} =\frac{1}{g'} B(r) \partial_{\mu}t 
-\frac{1}{g'}(1-\cos\theta) \partial_{\mu} \varphi.  
\label{ans}
\end{gather}
The ansatz has an apparent string singularity along the negative 
$z$-axis in $\xi$ and $B_{\mu}$. But they are a pure gauge 
artifact which can easily be removed making $U(1)_Y$ non-trivial. 
So the above ansatz describes a most general spherically symmetric 
ansatz of the electroweak dyon.

But this, of course, requires $U(1)_Y$ to be non-trivial. 
In other words, we need $U(1)_Y$ to be non-trivial to have 
the monopole. So the important question is why $U(1)_Y$ 
must be non-trivial. 

To understand this choose the unitary 
gauge with the gauge transformation 
\begin{gather}
\xi \rightarrow U \xi = \left(\begin{array}{cc} 0 \\ 1
\end{array} \right),  \nn\\
\vec A_\mu \rightarrow \frac{1}{g} \left( \begin{array}{c}
-f(r)(\sin\varphi\partial_\mu\theta
+\sin\theta\cos\varphi \partial_\mu\varphi) \\
f(r)(\cos\varphi\partial_\mu \theta
-\sin\theta\sin\varphi\partial_\mu\varphi) \\
A(r)\partial_\mu t -(1-\cos\theta)\partial_\mu\varphi
\end{array} \right),  
\end{gather}
and express the electromagnetic and $Z$-boson potentials 
$A_\mu^{\rm (em)}$ and $Z_\mu$ by
\begin{gather}
\left( \begin{array}{cc} A_\mu^{\rm (em)} \\ Z_{\mu}
\end{array} \right)
= \left(\begin{array}{cc}
\cos\theta_{\rm w} & \sin\theta_{\rm w}\\
-\sin\theta_{\rm w} & \cos\theta_{\rm w}
\end{array} \right)
\left( \begin{array}{cc} B_{\mu} \\ A^3_{\mu}
\end{array} \right) \nn\\
= \frac{1}{\sqrt{g^2 + g'^2}} \left(\begin{array}{cc} 
g & g' \\ -g' & g \end{array} \right)
\left( \begin{array}{cc} B_{\mu} \\ A^3_{\mu}
\end{array} \right), 
\label{unitary}
\end{gather}
where $\theta_{\rm w}$ is the Weinberg angle. Clearly 
$A_\mu^3$ has the string singularity along the negative 
z-axis. This is because the $U(1)$ subgroup of $SU(2)$ 
is non-trivial. This justifies the string singluarity 
in $B_\mu$, because $A_\mu^3$ already has the singularity. 
In other words, it is inconsistent (i.e., in violation 
of self-consistency) to insist $U(1)_Y$ to be trivial. 
This tells that the standard model must have the monopole. 

With (\ref{unitary}) we can express the Lagrangian (\ref{lag1}) 
in terms of the physical fields
\begin{gather}
{\cal L}= -\frac{1}{2}(\partial_\mu \rho)^2 
-\frac{\lambda}{8}\big(\rho^2-\dfrac{2\mu^2}{\lambda}\big)^2
-\frac{1}{4} F_{\mu\nu}^{\rm (em)2} 
-\frac{1}{4} Z_{\mu\nu}^2 \nn\\
-\frac{1}{2}|(D_\mu^{\rm (em)} W_\nu-D_\nu^{\rm (em)} W_\mu)
+ ie \dfrac{g}{g'} (Z_\mu W_\nu - Z_\nu W_\mu)|^2  \nn\\
-\frac{g^2+g'^2}{8} \rho^2 Z_\mu^2
-\frac{g^2}{4}\rho^2 |W_\mu|^2
+ie \frac{g}{g'} Z_{\mu\nu} W_\mu^* W_\nu  \nn\\
+ie F_{\mu\nu}^{\rm (em)} W_\mu^* W_\nu
+ \frac{g^2}{4}(W_\mu^* W_\nu - W_\nu^* W_\mu)^2,
\label{lag2}
\end{gather}
where $\rho$ and $W_\mu$ are the Higgs and $W$-boson, 
$D_\mu^{\rm (em)}=\partial_\mu+ieA_\mu^{\rm (em)}$, and 
$e=gg'/\sqrt{g^2+g'^2}$ is the electric charge. 

Moreover, the ansatz (\ref{ans}) becomes  
\begin{gather}
\rho =\rho(r),
~~~W_\mu= \frac{i}{g}\frac{f(r)}{\sqrt2}e^{i\varphi}
(\partial_\mu \theta +i \sin\theta \partial_\mu \varphi),
\nn\\
A_\mu^{\rm (em)}=e\big(\frac{A(r)}{g^2}+\frac{B(r)}{g'^2} \big)
\partial_\mu t -\frac1e (1-\cos\theta)\partial_\mu \varphi,  \nn\\
Z_\mu= \frac{e}{gg'}\big(A(r)-B(r) \big)\partial_\mu t.
\label{ansatz2}
\end{gather}
With this we have the equations of motion
\begin{gather}
\ddot f - \frac{f^2 -1}{r^2} f 
=\big(\frac{g^2}{4}\rho^2-A^2 \big) f,  \nonumber \\
\ddot \rho+\frac{2}{r}\dot\rho -\frac{f^2}{2r^2}\rho
=-\frac{1}{4}(B-A)^2 \rho +\lambda \big(\frac{\rho^2}{2}
-\frac{\mu^2}{\lambda}\big)\rho, \nonumber \\
\ddot A +\frac{2}{r} \dot A - \frac{2f^2}{r^2} A 
=\frac{g^2}{4}(A-B)\rho^2,  \nn\\
\ddot B+\frac{2}{r}\dot B =\frac{g'^2}{4}(B-A) \rho^2,
\label{eq1} 
\end{gather} 
which has a singular monopole solution
\begin{gather}
f=0,\quad\rho=\rho_0 = \sqrt{2\mu^2/\lambda},  
\nn\\
A_\mu^{\rm (em)} = -\frac{1}{e}(1- 
\cos \theta) \partial_\mu \varphi,~~~~Z_\mu=0.
\label{cmon}
\end{gather}
Choosing the boundary condition
\bea
&\rho(0)=0,~~f(0)=1,~~A(0)=0,~~B(0)=b_0, \nn\\
&\rho(\infty)=\rho_0,~f(\infty)=0,~A(\infty)=B(\infty)=A_0,
\label{bc}
\eea
we obtain the Cho-Maison dyon \cite{plb97}. 

We can also have the anti-monopole or in general anti-dyon 
solution, the charge conjugate state of the dyon, which 
has the magnetic charge $q_m=-4\pi/e$ with the following 
ansatz 
\begin{gather}
\phi' =\rho(r)~\xi',  
~~~\xi'=-i\left(\begin{array}{cc} \sin (\theta/2)~e^{+i\varphi}\\
- \cos(\theta/2) \end{array} \right),   \nn\\
\vec A'_{\mu}= -\frac{1}{g} A(r)\partial_{\mu}t~\hat r'
+\frac{1}{g}(f(r)-1)~\hat r' \times \pro_{\mu} \hat r', \nn\\
\hat r'=-\xi'^{\dagger} \vec \tau ~\xi',  \nn\\
B'_{\mu} =-\frac{1}{g'} B(r) \partial_{\mu}t 
+\frac{1}{g'}(1-\cos\theta) \partial_{\mu} \varphi.
\label{antid1}
\end{gather}
In terms of the physical fields the ansatz 
is expressed by 
\bea
&W'_{\mu}=\dfrac{i}{g}\dfrac{f(r)}{\sqrt2}e^{-i\varphi}
(\partial_\mu \theta -i \sin\theta \partial_\mu \varphi)
=-W_{\mu}^*, \nn\\
&A_{\mu}^{ \rm (em)} = -e\big( \dfrac{1}{g^2}A(r)
+ \dfrac{1}{g'^2} B(r) \big) \partial_{\mu}t  \nn\\
&+\dfrac{1}{e}(1-\cos\theta) \partial_{\mu} \varphi,  \nn \\
&Z'_{\mu} = -\dfrac{e}{gg'}\big(A(r)-B(r)\big) \partial_{\mu}t
=-Z_\mu.
\label{antid2}
\eea
This clearly shows that the the electric and magnetic charges
of the ansatz (\ref{antid1}) are the opposite to the dyon, 
which confirms that the ansatz indeed describes the anti-dyon.
Notice that the ansatz is basically the complex conjugation 
of the dyon ansatz. With this we obtain exactly the same equation 
of motion for the anti-dyon. 

The electroweak dyon has two remarkable features. First, 
unlike the Dirac monopole it has the magnetic charge $4\pi/e$ 
(not $2\pi/e$). Second, with a non-trivial dressing of weak 
bosons it looks similar to the Julia-Zee dyon \cite{julia}. 
But, unlike the Julia-Zee dyon, it has the point singularity 
at the origin \cite{plb97}.

The point singularity of the Cho-Maison monopole makes it 
difficult to estimate the mass classically. In the following
we predicts the monopole mass to be around 10 TeV, and present 
three ways, the dimensional argument, the scaling argument, 
and the ultraviolet regularization, to back up this.  \\

\noindent{\bf A. Dimensional Argument}

To estimate the order of the monopole mass, it is important
to realize that (roughly speaking) the monopole mass comes 
from the Higgs mechanism which generates the W-boson mass. 
To see this we first consider the 'tHooft-Polyakov monopole.
Let $\vec \Phi$ and $\A_\mu$ be the Higgs triplet and 
the gauge potential, and express the monopole ansatz by
\bea
&\vec \Phi= \rho~\hat r,
~~~\A_\mu=\C_\mu+\W_\mu,  \nn\\
&\C_\mu=-\dfrac1g \hr \times \pro_\mu \hr,
~~~\W_\mu=-f \C_\mu,
\eea 
where $\C_\mu$ is the Wu-Yang monopole potential. Notice that, 
exeept the overall amplitude $f$, the W-boson part of the ansatz
is given by the Wu-Yang potential. 

With this we have
\begin{gather}
|D_\mu \vec \Phi|^2=\frac12 (\pro_\mu \rho)^2
+\frac14 g^2 \rho^2 f^2 (\vec C_\mu)^2. 
\end{gather}
This (with $f \simeq 1$) tells that the monopole acquires mass 
through the Higgs mechanism, except that $\C_\mu$ contains 
the extra factor $1/g$. 

Similar mechanism works for the Weinberg-Salam model. Indeed 
with the ansatz (\ref{ans}) we have (with $A=B=0$)
\begin{gather}
|{\cal D}_\mu \phi|^2=\frac12 (\pro_\mu \rho)^2
+\frac12 \rho^2 |{\cal D}_\mu \xi|^2 \nn\\
=\frac12 (\pro_\mu \rho)^2+\frac18 g^2 \rho^2 f^2 (\vec C_\mu)^2. 
\end{gather}
So we could say that the Higgs mechanism is responsible for 
the mass of the electroweak monopole (more precisely 
the SU(2) part of it).

With this understanding, we can use the dimensional argument 
to predict the monopole energy. Since the monopole mass term 
in the Lagrangian contributes to the monopole energy we expect
\bea
E \simeq C \times \dfrac{4\pi}{e^2} M_W,
~~~C\simeq 1.
\label{mmass}
\eea 
This implies that the monopole mass should be about $1/\alpha$ 
times bigger than the electroweak scale, around 10 TeV. Now we 
have to know how to estimate $C$, and we discuss two ways to do 
so.  \\ 

\noindent{\bf B. Scaling Argument}

Suppose that the quantum correction removes the singularity 
at the origin. In this case we can use the Derrick's 
scaling argument to estimate the monopole mass, because the 
regularized solution should be stable under the rescaling of its 
field configuration. So consider the monopole configuration 
(with $A=B=0$) and let
\bea
&K_\phi=\Int |{\cal D}_i \phi|^2 d^3 x,
~V_\phi= \dfrac{\lambda}{2} \Int\big(|\phi|^2
-\dfrac{\mu^2}{\lambda} \big)^2 d^3x,  \nn\\
&K_A =  \dfrac{1}{4} \Int\vec{F}_{ij}^2 d^3x,
~~~K_B=\dfrac{1}{4} \Int B_{ij}^2 d^3x. 
\eea
Notice that $K_B$ makes the monopole energy infinite. 

Now, under the scale transformation $\vec x \rightarrow \lambda \vec x$,
we have  
\bea
&K_A \rightarrow \lambda K_A,~~~K_B \rightarrow \lambda K_B, \nn\\
&K_\phi \rightarrow\lambda^{-1} K_\phi,  
~~~V_\phi \rightarrow \lambda^{-3} V_\phi.
\eea
So we have the following condition for the
regularized monopole configuration
\bea
K_A + K_B = K_\phi + 3V_\phi.
\label{derrick}
\eea
Numerically we have  
\bea
&K_A \simeq 0.1852 \times\dfrac{4\pi}{e^2}{M_W},
~~~~K_\phi \simeq 0.1577 \times\dfrac{4\pi}{e^2}{M_W},  \nn \\
&V_\phi \simeq 0.0011 \times\dfrac{4\pi}{e^2}{M_W},
\eea
so that from (\ref{derrick}) we have
\bea
K_B \simeq 0.0058 \times \dfrac{4\pi}{e^2} M_W.
\eea
From this we can estimate the energy of the monopole 
\bea
E \simeq 0.3498 \times \dfrac{4\pi}{e^2} M_W \simeq 3.85~{\rm TeV}.
\eea
This strongly back up the dimensional argument. But here we have 
assumed the existence of a regularized monopole solution. Now we 
show how the quantum correction could regularize the 
singularity at the origin.  \\

\noindent{\bf C. Ultra-violet Regularization}

Notice that (\ref{lag2}) describes the ``bare" theory which should 
change to an ``effective" theory after the quantum correction
which changes the coupling constants to the scale dependent 
running couplings. To see how this quantum correction can make 
the monopole energy finite, let us consider the following 
effective Lagrangian 
\begin{gather}
{\cal L}_{eff} = -|{\cal D} _\mu \phi|^2
-\frac{\lambda}{2} \Big(\phi^2 -\frac{\mu^2}{\lambda}\Big)^2 
-\frac{1}{4} \vec F_\mn^2  \nn\\ 
-\frac{1}{4} \epsilon(|\phi|^2 ) G_\mn^2.
\label{effl}
\end{gather}
This type of effective Lagrangian has been used in non-linear 
electrodynamics and cosmology, and naturally appears in 
higher-dimensional unified theory \cite{prd87,prl92,babi}. 

Clearly $\epsilon$ effectively modifies $g'$ of the $U(1)_Y$ 
gauge coupling to $g' /\sqrt{\epsilon}$, but the Lagrangian 
still retains the $SU(2)\times U(1)_Y$ gauge symmetry. Moreover, 
when $\epsilon \rightarrow 1$ asymptotically, it reproduces 
the standard model. 

From (\ref{effl}) we have
\begin{gather}
\ddot{\rho} + \frac{2}{r}\dot{\rho}-\frac{f^2}{2r^2}\rho 
=-\frac{1}{4} (A-B)^2 \rho 
+\frac{\lambda}{2} (\rho^2- \rho_0^2) \rho \nn\\
+ \frac{\epsilon'}{g'^2}\Big(\frac{1}{r^4}-\dot{B}^2 \Big) \rho,  \nn\\
\ddot{f}-\frac{f^2-1}{r^2}f=\big(\frac{g^2}{4}\rho^2
- A^2\big)f, \nn\\
\ddot{A}+\frac{2}{r}\dot{A}-\frac{2f^2}{r^2}A
=\frac{g^2}{4}\rho^2(A-B), \nn \\
\ddot{B} + 2\big(\frac{1}{r}+
\frac{\epsilon'}{\epsilon} \rho \dot{\rho} \big) \dot{B}  
=-\frac{g'^2}{4 \epsilon} \rho^2 (A-B).
\end{gather}
This confirms that effectively $\epsilon$ changes the $g'$ 
to the ``running" coupling $\bar g'=g' /\sqrt{\epsilon}$. 
So, by making $\bar g'$ infinite at the origin, we can 
regularize the Cho-Maison monopole. 

\begin{figure}
\includegraphics[height=4cm, width=7cm]{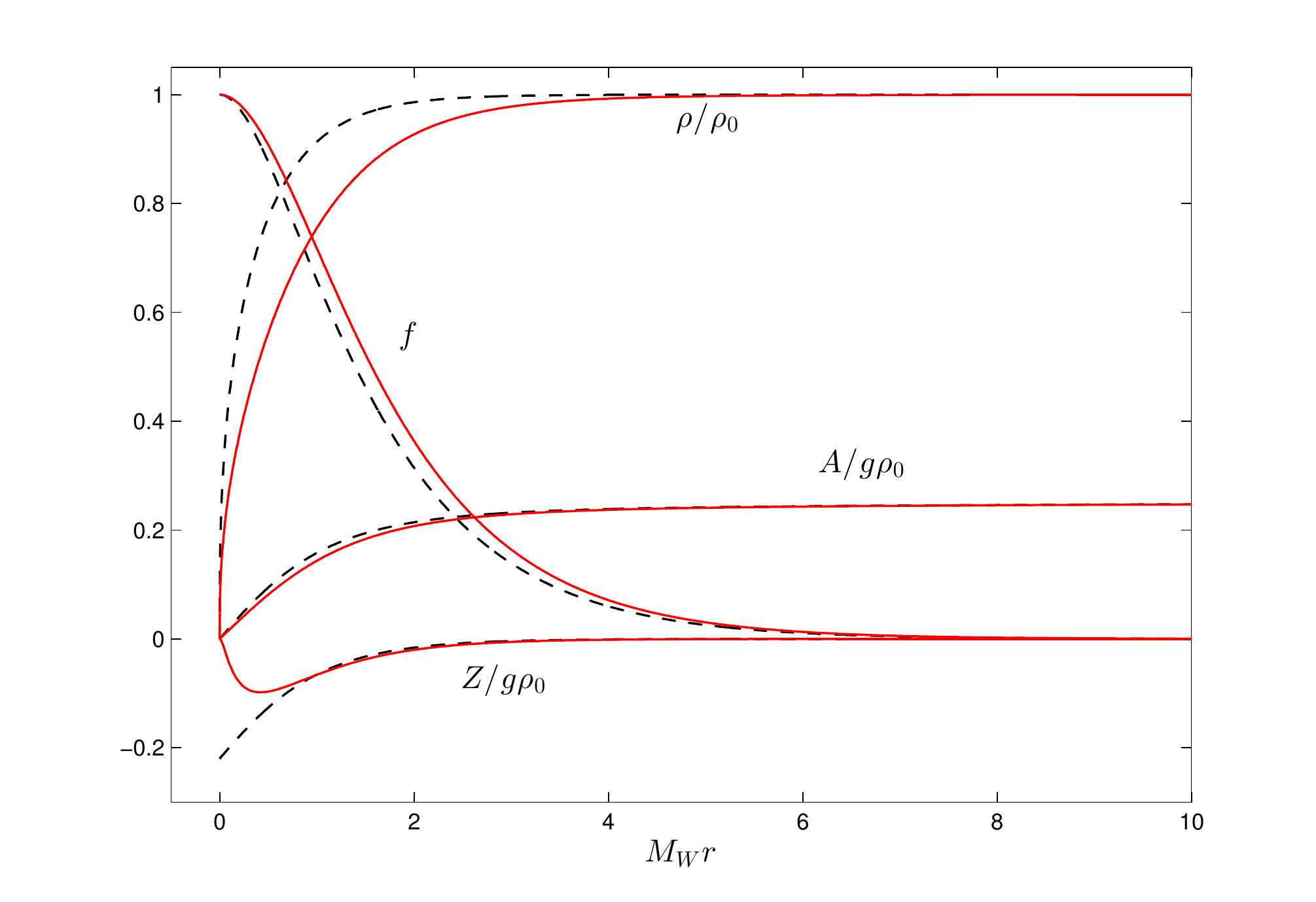}
\caption{\label{fcdyon} The electroweak dyon solution. 
The dotted and solid lines represent the Cho-Maison dyon
and the finite energy dyon ontained with the effective 
Lagrangian (\ref{effl}), where $Z=A-B$ and we have put 
$A(\infty)=M_W/2$.}
\label{fcdyon}
\end{figure}

Choosing $\epsilon = (\rho/\rho_0)^8$, we have the finite 
energy dyon solution shown in Fig. \ref{fcdyon} \cite{cho}. 
It is really remarkable that the regularized solution looks 
very much like the Cho-Maison dyon, except that for the 
finite energy dyon solution $Z(0)$ becomes zero. Moreover, 
with $A=B=0$ we can estimate the monopole energy 
\begin{gather}
E \simeq 0.65 \times \frac{4\pi}{e^2} M_W \simeq 7.19 ~{\rm TeV}.
\end{gather}
This tells two things. First, a quantum correction could easily 
make the electroweak monopole mass finite. Second, this strongly 
supports our estimate of the monopole mass based on the scaling 
argument.

We can think of another way to regularize the Cho-Maison monopole.
Suppose the quantum correction modifies (\ref{lag2}) by
\begin{gather}
\delta {\cal L}=ie \alpha F_{\mu\nu}^{\rm (em)} W_\mu^* W_\nu 
+\beta \frac{g^2}{4}(W_\mu^*W_\nu-W_\nu^*W_\mu)^2,
\label{modl}
\end{gather}
where $\alpha$ and $\beta$ are the scale dependent parameters 
which vanish asymptotically and modify the theory only at short 
distance. 

\begin{figure}
\includegraphics[height=4cm, width=7cm]{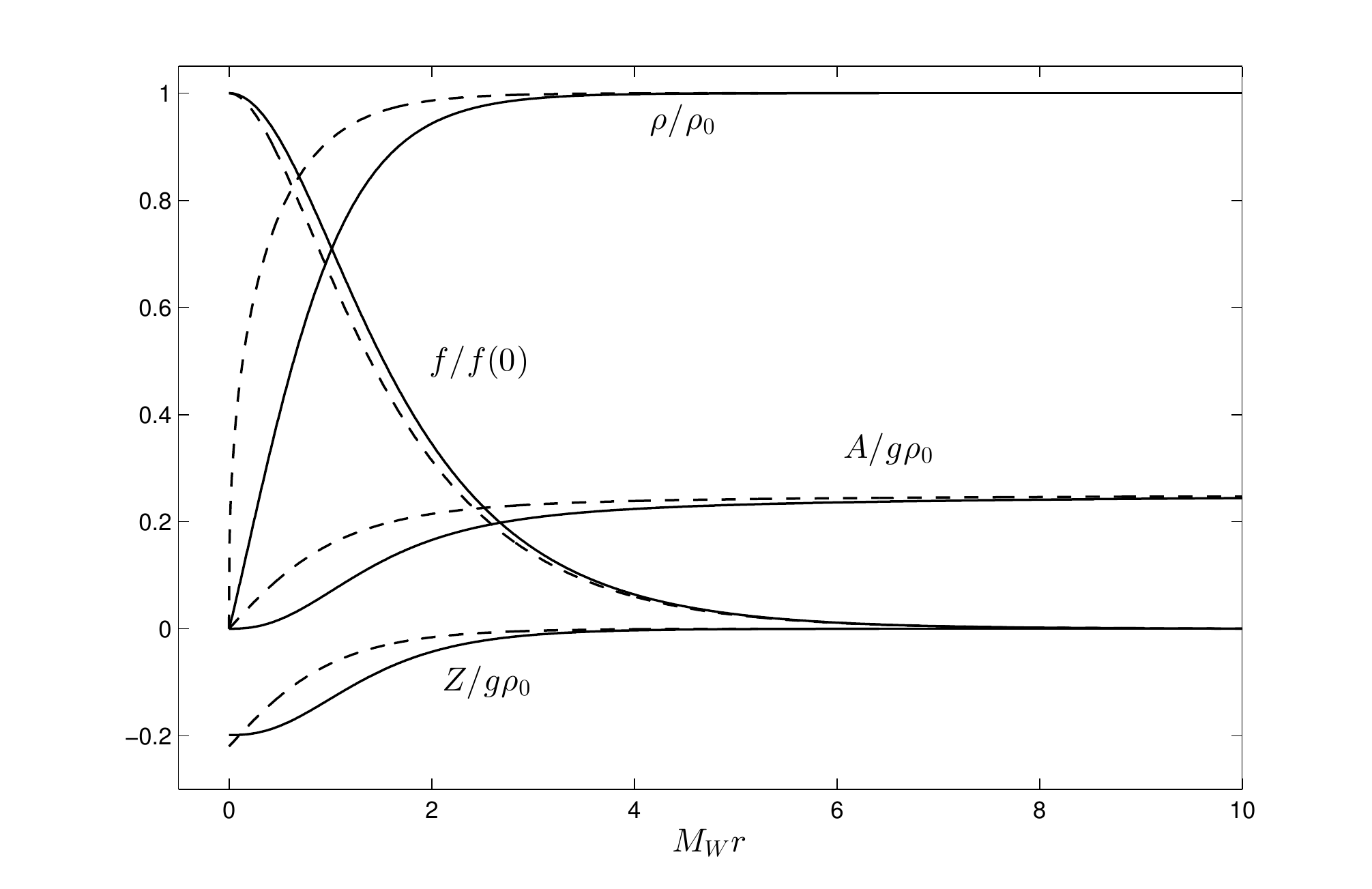}
\caption{\label{fecdyon} The finite energy electroweak dyon 
solution obtained from the modified Lagrangian (\ref{modl}). 
The solid line represents the finite energy dyon and dotted 
line represents the Cho-Maison dyon.}
\label{fig2}
\end{figure}

To find the finite energy dyon, however, we may treat $\alpha$ 
and $\beta$ as constants because asymtotically the boundary 
condition makes them irrelevant \cite{cho}. In this case the 
finite energy condition requires 
\begin{gather}
1+\alpha=\dfrac1{f(0)^2} \dfrac{g^2}{e^2},
~~~1+\beta=\dfrac1{f(0)^4} \dfrac{g^2}{e^2}.
\label{fecon}
\end{gather}
With this we can find a finite energy dyon solution with the 
boundary condition (\ref{bc}) but without the condition $f(0)=1$.
For instance, when $f(0)=g/e$ (with $\alpha=0$) we have the 
solution shown in Fig. \ref{fecdyon}. Notice that asymptotically 
boundary condition makes the solution converge to the Cho-Maison 
dyon. 

\begin{figure}
\includegraphics[height=4cm, width=8cm]{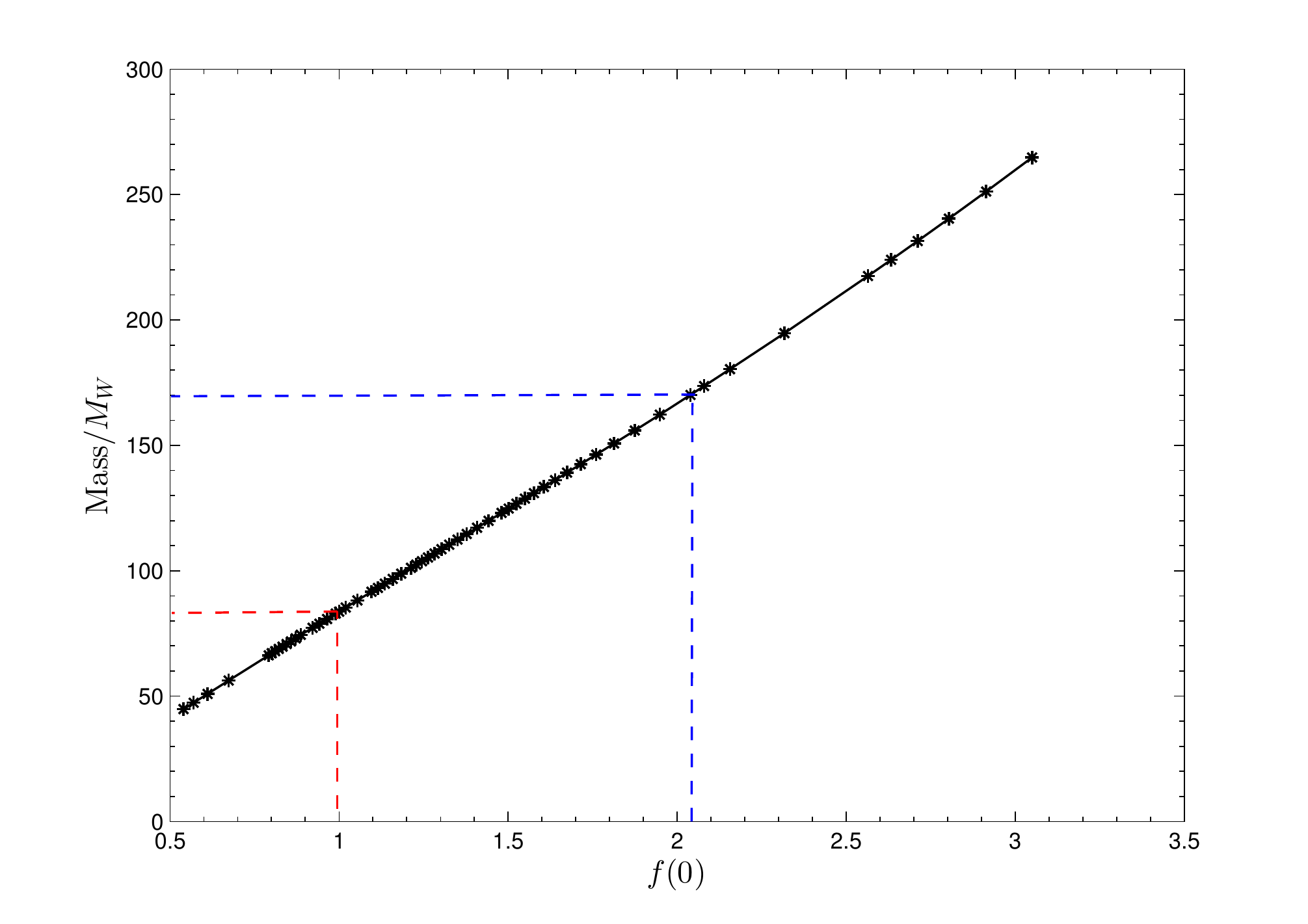}
\caption{\label{edf0} The energy dependence of the electroweak 
monopole on $f(0)$.}
\label{fig3}
\end{figure}

We can calculate the monopole energy in terms of $f(0)$. For 
example we have
\begin{eqnarray}
&E(f(0)=1) \simeq 0.61 \times \dfrac{4\pi }{e^2} M_W 
\simeq 6.73~{\rm TeV},  \nn\\
&E(f(0)=\dfrac{g}{e})\simeq 1.27 \times \dfrac{4\pi }{e^2} M_W 
\simeq 13.95~{\rm TeV}.
\end{eqnarray}
In general, we can plot the monopole energy as a function of 
$f(0)$, which is shown in Fig. \ref{edf0}. Again this assures 
that a simple quantum cotrrection could regularize the Cho-Maison
monopole, and strongly supports the our prediction 
of the monopole mass based on the scaling argument.

Of course (\ref{effl}) or (\ref{modl}) may not describe the true 
quantum correction, so that the finite energy solutions can only 
be viewed as approximate solutions of the standard model. But 
this is not the point. Our point here is to show that a small 
quantum correction, without new interaction, can regularize 
the Cho-Maison dyon. As Fig. 1 and Fig. 2 demonstrate, they 
describe an excellent approximation of the Cho-Maison dyon from 
which we can estimate the mass of the electroweak monopole.  

We close with the following remarks:  \\
1. In electrodynamics $U(1)$ can either be trivial or 
non-trivial. But in the standard model it is natural to 
expect $U(1)_{(\rm em)}$ to be non-trivial. As we have 
emphasized, there is no reason why $U(1)_Y$ has to be 
trivial since the $U(1)$ subgroup of $SU(2)$ is already 
non-trivial. This assures $U(1)_{(\rm em)}$ made of the 
linear combination of them to be non-trivial. This means 
that, unlike the Dirac monopole in electrodynamics which 
is optional, the standard model must have the monopole. \\
2. The unit magnetic charge of the electroweak monopole must 
be $4\pi/e$. This again is because $U(1)_{(\rm em)}$ is 
made of the linear combination of $U(1)_Y$ and $U(1)$ subgroup 
of $SU(2)$, which makes the period of $U(1)_{(\rm em)}$ 
$4\pi$.    \\
3. Since the monopole singularity comes from $B_\mu$, we can 
regularize it embedding the hypercharge $U(1)$ to $SU(2)_Y$. 
This, of course, will make the monopole mass heavier because 
this embedding adds an intermediate scale $M_Y$ (somewhere 
between the electroweak and grand unification scales) \cite{cho}. 
Such embedding could naturally arise in the left-right symmetric 
grand unification models, in particular in the $SO(10)$ grand 
unification. 

Certainly the existence of the electroweak monopole of 
mass around 4 to 10 TeV has important implications. 
First, this explains why the search for the monopole 
so far has been unsuccessful. Moreover, this implies that
the recent upgrading of LHC could allow the MoEDAL to 
detect it, because the LHC nopw might have reached 
the monopole-antimonopole pair production threshold. 
But most importantly, it tells that the final test 
of the standard model should be the discovery of the 
electroweak monopole, not the Higgs particle. Indeed 
this should be regarded as the topological test of 
the standard model, which has never been done before. 

But if the monopole mass becomes above 7 TeV, the 14 TeV
LHC can not produce it. In this case we may need the 
``cosmic" MoEDAL to detect it, or the 100 TeV LHC. A detailed 
discussion of our work will be published in a separate 
paper \cite{cho}.

\noindent{\bf Acknowledgments}

The work is supported in part by the National Research Foundation 
of Korea funded by the Ministry of Education, Science, and 
Technology (2012-002-134).

\end{document}